\documentclass[a4paper,11pt]{article}
\usepackage{pos}
\usepackage{amsmath}
\usepackage{subcaption}
\usepackage{epsfig}
\usepackage{color} 
\usepackage{soul}

 \usepackage{multirow}
 \usepackage{enumitem}
\usepackage{amsfonts}
\usepackage[output-decimal-marker={.}]{siunitx}

\title{Tetraquark channels with $\bar b b$ pair in the static limit}

\author*[a]{Mitja Sadl}
\author[a,b]{Sasa Prelovsek}

\affiliation[a]{Faculty of Mathematics and Physics, University of Ljubljana,\\
  1000 Ljubljana, Slovenia}

\affiliation[b]{Jozef Stefan Institute,\\
1000 Ljubljana, Slovenia}

\emailAdd{mitja.sadl@fmf.uni-lj.si}
\emailAdd{sasa.prelovsek@ijs.si}

\abstract{Belle experiment discovered two hadrons with exotic quark content $Z_b^+\simeq \bar bb \bar du$. We present a lattice study of the $\bar bb\bar du$ systems with various quantum numbers using static bottom quarks. Only one set of quantum numbers that couples to $Z_b$ and $\Upsilon\;\pi$ was explored on the lattice before: these studies found an attractive potential between $B$ and $\bar B^*$ which leads to a bound state below the threshold. In the present study, we consider the other three sets of quantum numbers. Eigen-energies of the $\bar bb \bar du$ system are extracted as a function of separation between $b$ and $\bar b$. The resulting eigen-energies do not show any sizable deviation from non-interacting energies of the systems $\bar bb+\bar du$ and $\bar bu+\bar db$, so no significant attraction or repulsion is found.  A slight exception is a small attraction between  $B$ and $\bar B^*$ at small distance for the quantum number that couples to $Z_b$ and $\eta_b\;\rho$.}

\FullConference{%
 The 38th International Symposium on Lattice Field Theory, LATTICE2021
  26th-30th July, 2021
  Zoom/Gather@Massachusetts Institute of Technology
}

\begin{document}
\maketitle

\section{Introduction}\label{sec:intro}

The Belle experiment announced a discovery of two tetraquarks $Z_b(10610)$ and $Z_b(10650)$ with $J^{P}\!=\!1^+$ and $I\!=\!1$ in 2011 \cite{Belle:2011aa,Garmash:2014dhx}. Both resonances were first observed in decays to $Z_b^{\pm}\to \Upsilon (nS) \pi^{\pm}$ and $Z_b^{\pm}\to h_b (mP) \pi^{\pm}$, which indicates the exotic flavor content $Z_b^+\sim \bar{b}b\bar{d}u$. Later, Belle established that $Z_b(10610)$ and $Z_b(10650)$ predominantly decay to $B\bar{B}^*$ and $B^*\bar{B}^*$, respectively \cite{ParticleDataGroup:2020ssz}. Their masses are slightly above these two thresholds. Many phenomenological studies indicate that the $B^{(*)}\bar B^*$ molecular Fock component is essential for $Z_b$ (see for example \cite{Wang:2018jlv}). Furthermore, in \cite{Wang:2018jlv,Bondar:2011ev} $Z_b(10610)$ and $Z_b(10650)$ are dominated by $B\bar B^*$ and $B^*\bar B^*$, respectively,

No lattice studies of the $\bar{b}b\bar{q}q$ resonances via the rigorous L{\"u}scher formalism are available. This is too challenging at present since one would have to determine a scattering matrix of at least seven coupled channels from a very dense spectrum of eigen-energies.

\begin{figure*}[ht!]
\begin{tabular}{cccc} 
         \multirow{4}{*}{
	\begin{subfigure}{0.28\textwidth}
	\vspace{2.0cm}
	\includegraphics[width=\linewidth]{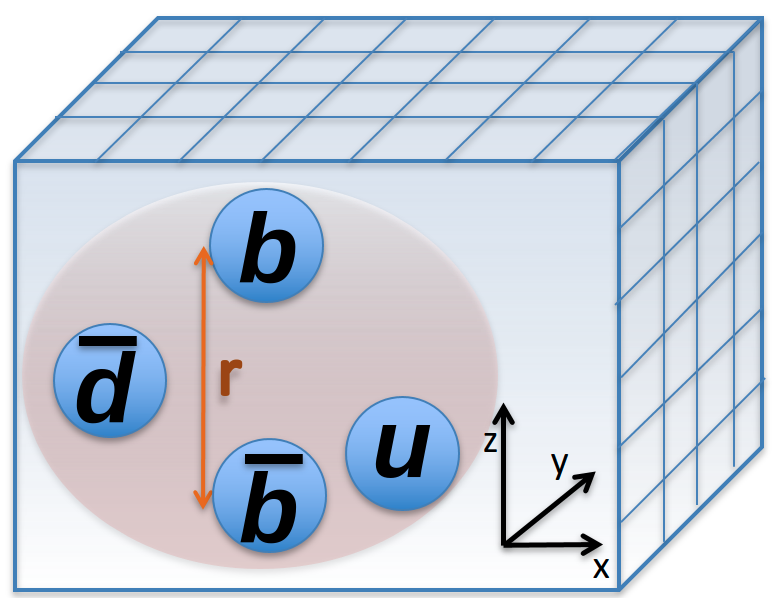}
	\vspace{1.95cm}
	\caption{ } \label{fig:1a}
	\end{subfigure}
								}         &   
	\begin{subfigure}{0.18\textwidth}
	\includegraphics[width=\linewidth]{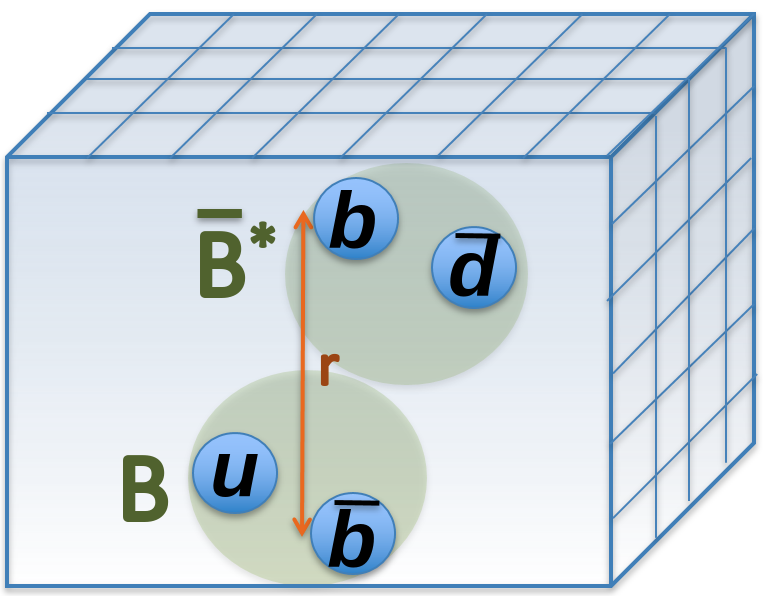}
	\end{subfigure}     
						  &   
	\begin{subfigure}{0.18\textwidth}
	\includegraphics[width=\linewidth]{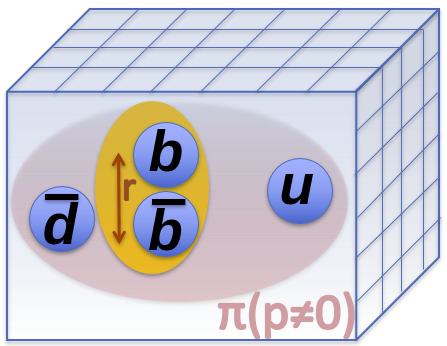}
	\end{subfigure}     
							   &   		   \\
                                     &   
	\begin{subfigure}{0.18\textwidth}
	\includegraphics[width=\linewidth]{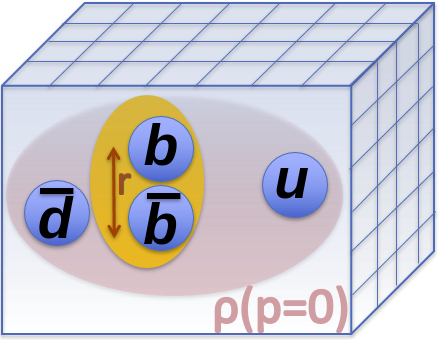}
	\end{subfigure}     
						  &   
	\begin{subfigure}{0.18\textwidth}
	\includegraphics[width=\linewidth]{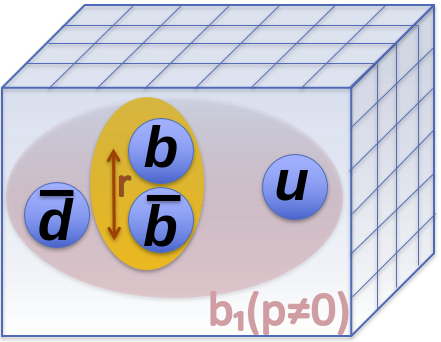}
	\end{subfigure}     
							   &    
	\begin{subfigure}{0.18\textwidth}
	\includegraphics[width=\linewidth]{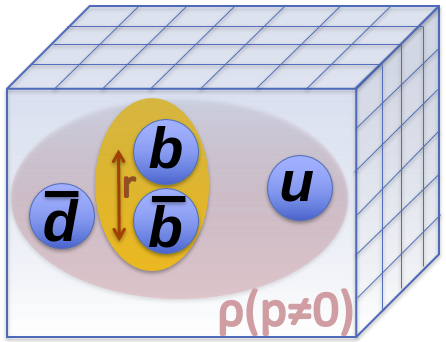}
	\end{subfigure}     
									   \\
                                     &   
	\begin{subfigure}{0.18\textwidth}
	\includegraphics[width=\linewidth]{4q_illustration_rhop.png}
	\end{subfigure}     
						  &   
	\begin{subfigure}{0.18\textwidth}
	\includegraphics[width=\linewidth]{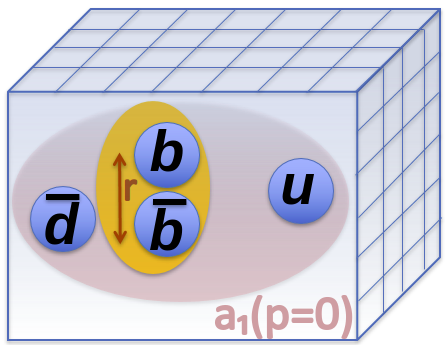}
	\end{subfigure}     
							   &    
	\begin{subfigure}{0.18\textwidth}
	\includegraphics[width=\linewidth]{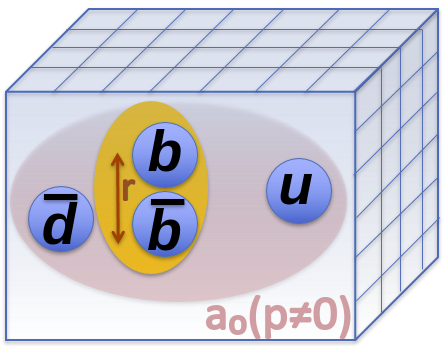}
	\end{subfigure}     
									   \\
                                     &   
	\begin{subfigure}{0.18\textwidth}
	\vspace{2.4cm}
	\caption{$J_z^l=0,\\ C\!\cdot\! P=+1,$ $\epsilon=+1$} \label{fig:1b}	
	\end{subfigure}     
						  &   
	\begin{subfigure}{0.18\textwidth}
	\vspace{-0.2cm}
	\includegraphics[width=\linewidth]{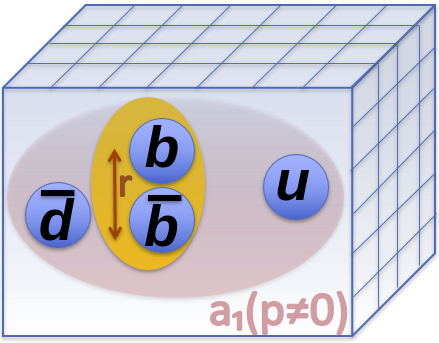}
	\caption{$J_z^l=0,\\ C\!\cdot\! P=+1,$ $\epsilon=-1$} \label{fig:1c}	
	\end{subfigure}     
							   &    	
	\begin{subfigure}{0.18\textwidth}
	\vspace{2.4cm}
	\caption{$J_z^l=0,\\ C\!\cdot\! P=-1,$ $\epsilon=+1$} \label{fig:1d}	
	\end{subfigure}     	   \\
\end{tabular}
\caption{(a) The system studied with static $b$ and $\bar b$; (b,c,d) the states of this system with various quantum numbers captured by our operators. The operators for the (b) are shown in \eqref{E3}, while the operator for (c) and (d) can be found in \cite{Sadl:2021bme}.}\label{fig:1}
\end{figure*}

This work considers the $\bar bb\bar du$ system with static $b$ and $\bar b$ quarks fixed to distance $r$ within lattice QCD, as shown in Fig. \ref{fig:1a}. The goal is to determine eigen-energies of this system $E_n(r)$ as a function of separation $r$ for various quantum numbers. The resulting energies are then compared to the non-interacting (n.i.) energies $E^{\textrm{n.i.}}(r)$ of subsystems $[\bar bb][\bar du]$ and $[\bar bu][\bar db]$, where $[..]$ denotes a color-singlet meson of a given flavor. The eigen-energies represent lattice input to study this system within the Born-Oppenheimer approximation. This can be done by solving the nonrelativistic Schr{\"o}dinger equation with the static potential according to the general strategy outlined, for example, in \cite{BO,Braaten:2014qka}.

The $Z_b$ with $J^P\!=\!1^+$ corresponds in the molecular $B^{(*)}\bar B^{*}$ picture to the linear combination of two quantum channels in the static limit
\begin{align}
\label{decomposition}
B \bar B^*_k + B^*_k\bar B
&\propto (S^h\!=\!0)(J^l\!=\!1,\  C\! \cdot\! P\!=\!\epsilon\!=\!+1)+(S^h\!=\! 1)(J^l\! =\! 0,\  C\!\cdot\!P\!=\!\epsilon\!=\!-1) \nonumber\\ 
B^*_i \bar B^*_j - B^*_j\bar B^*_i
& \propto (S^h\!=\!0)(J^l\!=\!1,\  C\! \cdot\! P\!=\!\epsilon\!=\!+1)- (S^h\!=\!1)(J^l\!=\!0,\  C\!\cdot\!P\!=\!\epsilon\!=\!-1)\>.
\end{align}
This can be rigorously shown with the Fierz transformations (see Eq. (3) in our longer publication \cite{Sadl:2021bme}). The total spin of heavy quarks ($S^h$) and the angular momentum of the light degrees of freedom ($J^l$) are separately conserved in the static limit $m_b\!\to\! \infty$. Let us for the moment postpone the explanation of the connection to $C\!\cdot\!P$ and $\epsilon$.

Lattice simulations of $Z_b$ \cite{Peters:2016wjm, Prelovsek:2019ywc} have been done only for the quantum number  $J^l\! =\! 0$, where $Z_b$ couples to $\Upsilon\;\pi$ and to the second component of  $B^{(*)}\bar B^*$ on the right-hand side of \eqref{decomposition}. Both available studies found that the eigenstate dominated by $B\bar{B}^*$ has energy significantly below $m_B+m_{B^*}$ at small $r$. This rendered the static potential with sizable attraction  between $B$ and $\bar B^*$ at small $r$. The nonrelativistic Schr{\"o}dinger equation for $B\bar{B}^*$ leads to a bound state below the $B\bar{B}^*$ threshold, which could be related to $Z_b$.

The present lattice study (see our publication \cite{Sadl:2021bme}) considers other three sets of quantum numbers (see \ref{fig:1b}, \ref{fig:1c}, \ref{fig:1d}) for the $\bar bb\bar du$ system. These quantum numbers have not been  studied before,  with exception of \cite{Alberti:2016dru}  that  considered  the ground state of one channel. We investigate the quantum number (\ref{fig:1b}) which contains $J^l\! =\!1$ and is relevant  for $Z_b$, where this resonance couples to $\eta_b\;\rho$ and to the first component of $B\bar B^*$ \eqref{decomposition}.

In addition, we study two other sets of quantum numbers (\ref{fig:1c}, \ref{fig:1d}) which do not couple to  $B\bar B^*$ but only to $[\bar bb][\bar du]$ in the explored energy region.

\section{Quantum numbers and operators}\label{sec:QNandOP}

In the static approximation $m_b\to \infty$, the conserved quantum numbers differ from those when $b$ and $\bar{b}$ have finite mass. In our case we consider the $\bar{b}b\bar{q}q$ system in Fig. \ref{fig:1a} and define the axis of separation of $b$ and $\bar b$ to be the $z$-axis. The good quantum numbers are thus isospin $I$, its third component $I_3$, angular momentum of the light degrees of freedom $J^{l}_z$, the product of parity and charge conjugation $C\!\cdot\! P$ and the reflection over $yz$-plane $\epsilon$. More details can be found in Sec. II of \cite{Sadl:2021bme}.

We study the four-quark system $\bar{b}b\bar{q}q$ with  
 \begin{equation}
\label{E2}
I=1,\ I_3=0, \  J^{l}_z=0~. 
\end{equation}
Table \ref{tab:1} lists the  three sets of quantum numbers considered here and one set considered  in the previous studies \cite{Peters:2016wjm, Prelovsek:2019ywc}.

\begin{table}[h!]
\centering
\hspace{-1.2cm}\begin{tabular}{ccccc|c|cc|c} 
\hline
\hline
\multicolumn{8}{c|}{quantum numbers}                                                                                & \multirow{2}{*}{lat.  studies}           \\ 
\cline{1-8}
 $I$                & $I_3$              & $J_z^{l}$ & $C\!\cdot\! P$ & $\epsilon$ & $\Lambda^{\epsilon}_{C\!P}$  & $S^h$ & $S^h_z$ &                             \\ 
\hline
 \multirow{4}{*}{\vspace{-0.3cm}1} & \multirow{4}{*}{\vspace{-0.3cm}0} & \multirow{4}{*}{\vspace{-0.3cm}0} & $-1$       & $-1$       & $\Sigma_u^-$       &  \multirow{4}{*}{\vspace{-0.3cm}0,1}  & \multirow{4}{*}{\vspace{-0.3cm}0}         & \cite{Prelovsek:2019ywc,Peters:2016wjm}                        \\ 
\cline{9-9}
                     &                    &                    & $+1$       & $+1$       & $\Sigma_g^+$   &&            & \multirow{3}{*}{this work \cite{Sadl:2021bme}}  \\
                   &                    &                    & $+1$       & $-1$       & $\Sigma_g^-$           &&    &                             \\
                   &                    &                    & $-1$       & $+1$       & $\Sigma_u^+$           &&    &                             \\
\hline
\hline
\end{tabular}
\caption{Four sets of quantum numbers for the system $\bar bb\bar qq$: The first one was studied in \cite{Prelovsek:2019ywc,Peters:2016wjm}, whereas we study the other three. The system is invariant under the rotations of the heavy quark spins, so  the results are independent of $S^h$. $\Lambda^{\epsilon}_{\eta=C\!P}$ is written according to the convention in \cite{Juge:1999ie}.}
\label{tab:1}
\end{table}

Our operators resemble Fock components $[\bar b q][\bar qb]$ and $[\bar b b][\bar qq]$, schematically shown in Fig. \ref{fig:1} with quantum numbers represented in Table \ref{tab:1}. Employed annihilation operators for the quantum numbers $J_z^l=0,\ C\!\cdot\! P= +1,\ \epsilon=+1$ are listed below (operators for the other two channels $J_z^l=0,\ C\!\cdot\! P= +1,\ \epsilon=-1$ and $J_z^l=0,\ C\!\cdot\! P= -1,\ \epsilon=+1$ are provided in \cite{Sadl:2021bme}): 
\vspace{-.7cm}
\begin{align}
\label{E3}
O_1\!&=\! O_{B\bar B^*}\!\!\propto\sum_{a,b}\sum_{A,B,C,D}\!\!\!\!\Gamma_{BA}\tilde \Gamma_{CD}~\bar b^a_C(0)q_A^a(0)~ \bar q^b_B(r)b_D^b(r) \nonumber \\ 
&\propto \bigl([\bar b(0) P_- \gamma_5 q(0)]~[\bar q(r) \gamma_z P_+ b(r)]+\{\gamma_5\leftrightarrow \gamma_z\}\bigr)   \nonumber \\ 
& \ - \bigl( [\bar b(0) P_- \gamma_y q(0)]~[\bar q(r) \gamma_x P_+ b(r)]+ \{\gamma_y\leftrightarrow \gamma_x\}\bigr)  \nonumber \\
O_2\!&=\! O_{(B\bar B^*)'} \nonumber\\
O_3\!&=\!O_{[\bar{b}b] \rho(0)}\!\propto\! [\bar b(0) U\Gamma^{\textrm{(H)}} b(r)]~[\bar q \gamma_zq]_{\vec p=\vec 0}\nonumber\\
O_4\!&=\!O_{[\bar{b}b] \rho(1)}\!\propto\! [\bar b(0) U\Gamma^{\textrm{(H)}} b(r)]~ \bigl([\bar q \gamma_zq]_{\vec p=\vec e_z}+ [\bar q \gamma_zq]_{\vec p=-\vec e_z}\bigr)\nonumber\\
O_5\!&=\!O_{[\bar{b}b] \rho(2)}\!\propto\! [\bar b(0) U\Gamma^{\textrm{(H)}} b(r)]~ \bigl([\bar q \gamma_zq]_{\vec p=2\vec e_z}+ [\bar q \gamma_zq]_{\vec p=-2\vec e_z}\bigr) ~.
\end{align}
Let us provide some comments on operators. For more details on the operators we refer the reader to \cite{Sadl:2021bme}. The gamma matrices sandwiched between static quarks can be $\tilde\Gamma,\Gamma^{\textrm{(H)}}=\gamma_5P_+$  or $ \gamma_zP_+$ for $S^h=0$ or $1$, respectively. The static limit implies that the correlators and $E_n$ are the same for both, so our results apply to both cases. The operators $O_{B\bar B^*}$ and $O_{(B\bar B^*)'}$ resembling $[\bar bq][\bar qb]$ are constructed with $\Gamma\!=\!P_-\gamma_z$ that satisfies $J^{l}_z=0$. Our operators for the channels $C\!\cdot\! P= +1,\ \epsilon=-1$ and $C\!\cdot\! P= -1,\ \epsilon=+1$ do not contain this kind of operators since these quantum numbers do not couple to a pair of negative parity $B$-mesons. The second and third line in \eqref{E3} are obtained via the Fierz transformation, where we take $\tilde \Gamma\!=\!\gamma_5 P_+$. This decomposition clarifies why this quantum channel is a linear superposition of $B\bar B^*$, $B^* \bar B$ and $B^* \bar B^*$ and why we labeled the two terms on the right-hand side of \eqref{decomposition} with $C\!\cdot\! P$ and $\epsilon$. Throughout this paper we refer to any combination of $B^{(*)}\bar B^{(*)}$ as $B\bar B^{*}$. One should namely note that in the static limit, $B$ and $B^*$ mesons are degenerate.

The operators resembling $[\bar bb][\bar qq]$ are formed from a color-singlet bottomonium and color-singlet light-meson current. The light current $[\bar q \Gamma^\prime q]$ with $\Gamma^\prime=\gamma_5$ (present in $C\!\cdot\! P= +1,\ \epsilon=-1$ and $C\!\cdot\! P=\epsilon=-1$ of \cite{Prelovsek:2019ywc}) couples in the low energy region to a pion, $l=\pi$. The currents for other $\Gamma^\prime$ couple to resonances $l=\rho,~b_1~a_1,~a_0$ that are not strongly stable on our lattice. So these currents in principle couple also to the allowed strong decay products of those resonances. The reliable and rigorous extraction of eigen-energies would require implementation of multi-hadron operators in the light sector which is beyond the scope of the present study. In practice, the employed operator $[\bar q \Gamma^\prime q]_{\vec p}$ couples to one finite-volume energy level $E_{l(\vec p)}$ which is a mixture of resonant and multi-hadron eigenstates in practice. Our main purpose is to find out whether there is some interaction between bottomonium $[\bar bb]$ and the light degrees of freedom $l$. Therefore we compare the sum of the separate energies $V_{\bar bb}+E_{l(\vec p)}$ with the eigen-energy $E_n$ of the whole system $[\bar bb][\bar qq]$, where the light degrees of freedom arise from the same current $[\bar q \Gamma^\prime q]_{\vec p}$ in both cases. This strategy does not lead to a complete spectrum of eigen-energies, but it still indicates whether the energy of light degrees of freedom is affected in the presence of a bottomonium.

\section{Lattice details}\label{sec:details}

Simulation is performed on an ensemble with dynamical Wilson-clover $u/d$ quarks,  $m_\pi \simeq \SI{266(5)}{MeV}$, $a\simeq \SI{0.1239(13)}{fm}$ and 281 configurations \cite{Hasenfratz:2008ce,Lang:2011mn}. We employ an ensemble with $N_L\!=\!16$ ($L\!\simeq\! \SI{2}{fm}$) and $N_T\!=\!32$. The latter is effectively doubled by summing the light-quark propagators with periodic and anti-periodic boundary conditions in time \cite{Lang:2011mn}. The Wick contractions are calculated using distillation, and energies are extracted via GEVP.

\section{Eigen-energies of  $\bar bb\bar du$ system as a function of $r$}\label{sec:eigen-energies}

\begin{figure*}[ht!]
\centering
\hspace*{-0.9cm}
\begin{subfigure}{0.46\textwidth}
\includegraphics[width=1.15\textwidth]{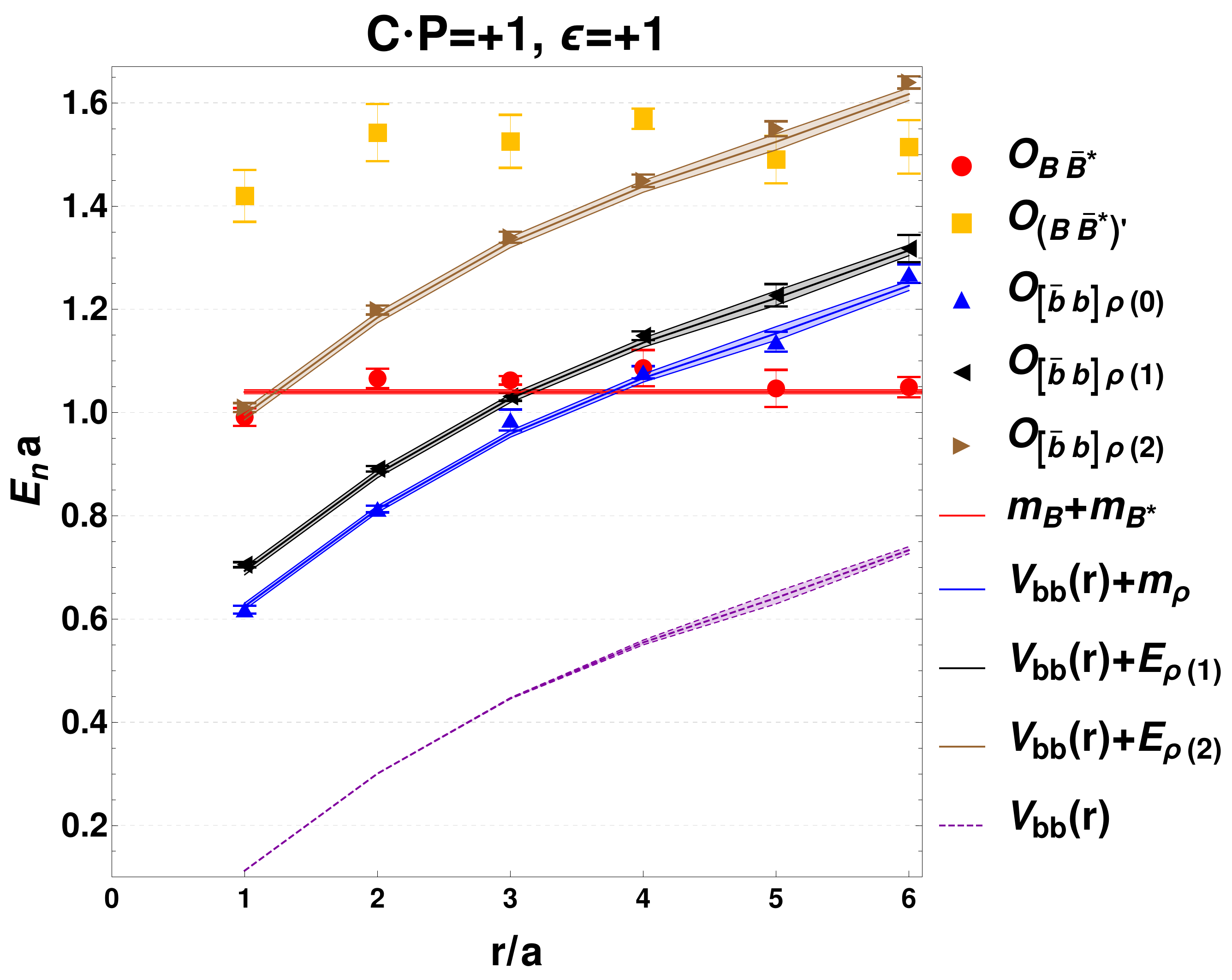}
\caption{$I=1,~J^l_z=0,~C\!\cdot\! P=\!+1, ~ \epsilon\!=\!+1$ } \label{fig:2a}
\end{subfigure} 
\hspace*{1cm}
\begin{subfigure}{0.46\textwidth}
\includegraphics[width=1.1\textwidth]{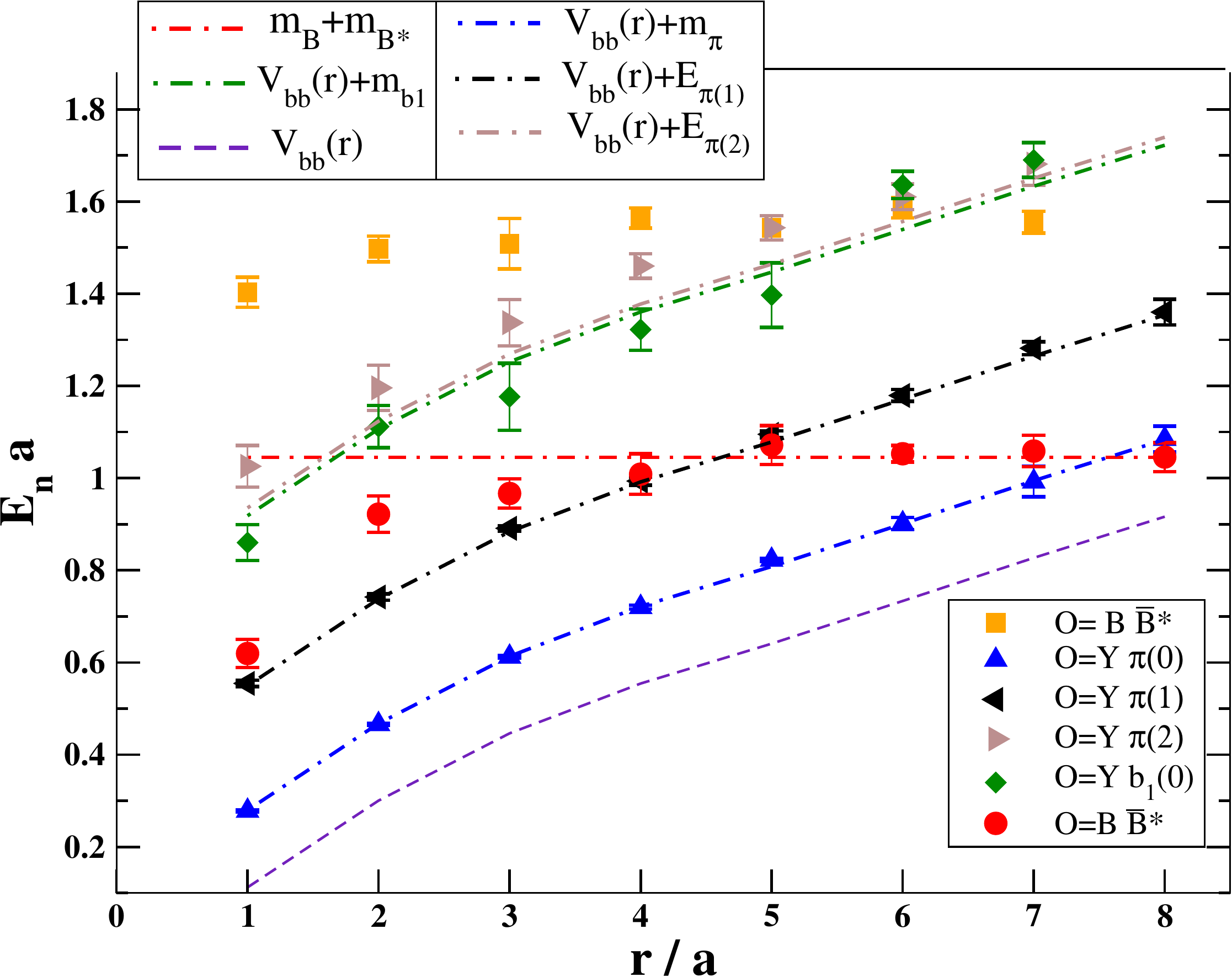}
\caption{ $I=1,~J^l_z=0,~C\!\cdot\! P=\!-1, ~ \epsilon\!=\!-1$} \label{fig:2b}
\end{subfigure}
\caption{Eigen-energies for the two channels of the system $\bar bb\bar qq$ that include the $[\bar bq][\bar qb]$ operators. On the left are the results for quantum numbers $C\!\cdot\! P\!=\!\epsilon\!=\!+1$. Eigen-energies are shown by symbols for separations between static quarks $b$ and $\bar b$. The labels indicate which two-hadron component dominates each eigenstate. The lines represent related two-hadron energies $E^{\textrm{n.\,i.}}$ \eqref{E6} when two hadrons do not interact. The width of their bands shows the uncertainty. On the right is a similar plot from \cite{Prelovsek:2019ywc} with the results for $C\!\cdot\! P\!=\!\epsilon\!=\!-1$. Lattice spacing is $a\simeq\SI{0.124}{fm}$ for both cases.}
\label{fig:2}
\end{figure*}

\begin{figure*}[ht!]
\centering
\hspace*{-0.9cm}
\begin{subfigure}{0.46\textwidth}
\includegraphics[width=1.15\textwidth]{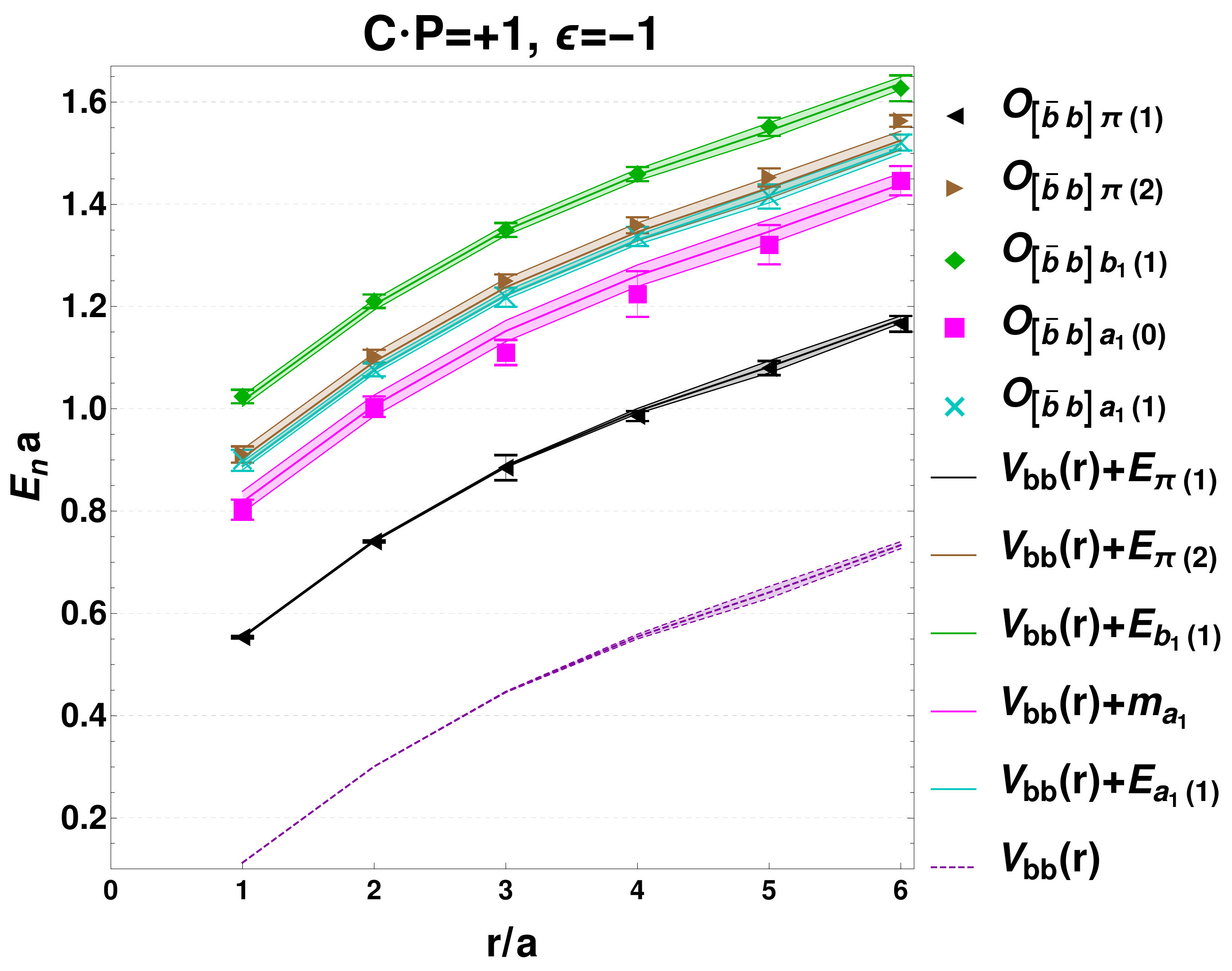}
\caption{$I=1,~J^l_z=0,~C\!\cdot\! P=\!+1, ~ \epsilon\!=\!-1$ } \label{fig:3a}
\end{subfigure} 
\hspace*{1cm}
\begin{subfigure}{0.46\textwidth}
\includegraphics[width=1.15\textwidth]{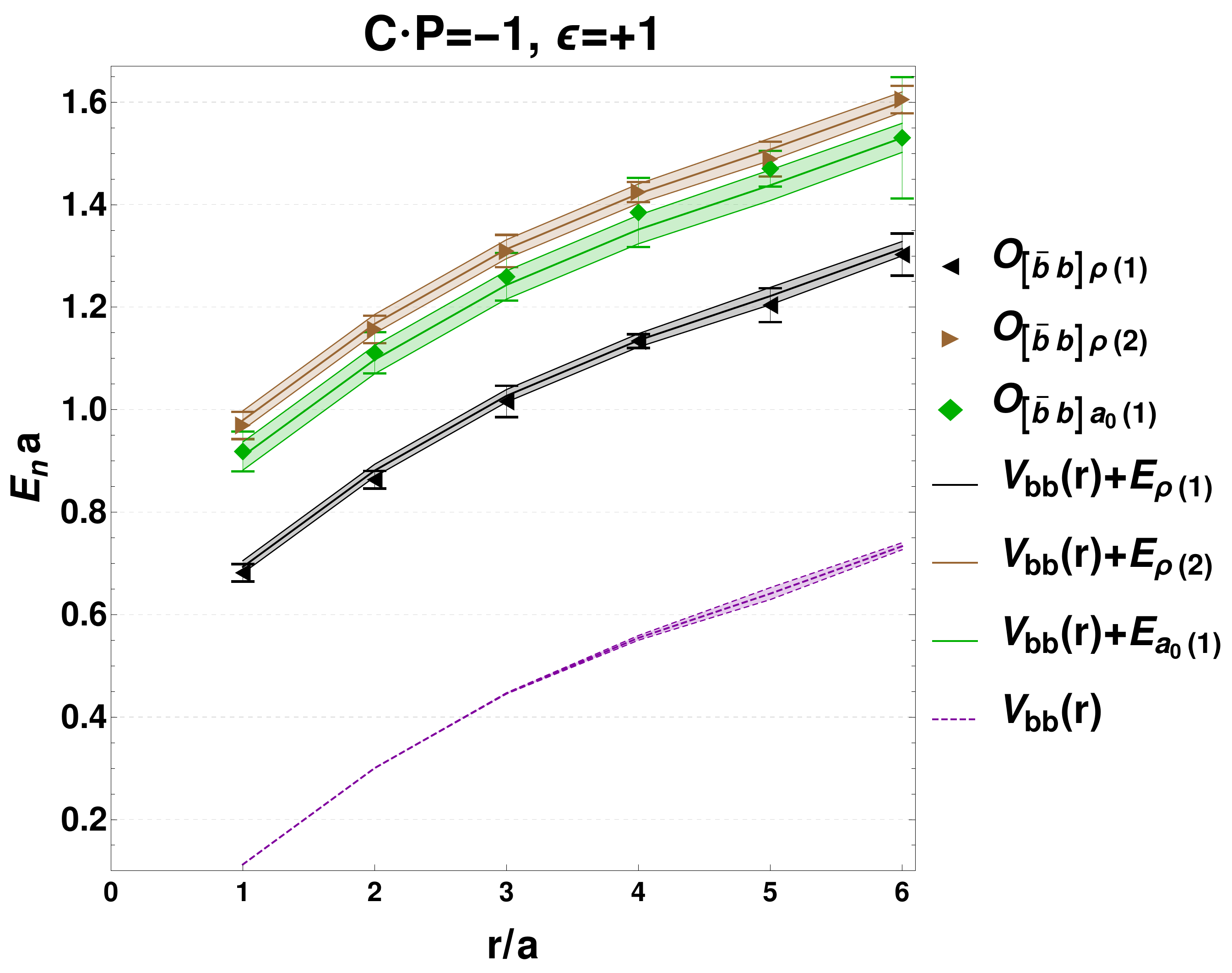}
\caption{ $I=1,~J^l_z=0,~C\!\cdot\! P=\!-1, ~ \epsilon\!=\!+1$} \label{fig:3b}
\end{subfigure}
\caption{The eigen-energies $E_n$ (symbols) and the two-hadron non-interacting energies $E^{\textrm{n.\,i.}}$ (lines) for the two remaining channels similarly as in Fig. \ref{fig:2}.}
\label{fig:3}
\end{figure*}

The central results of our study are the eigen-energies of the $\bar bb\bar du$ system (Fig. \ref{fig:1a}) with static $b$ and $\bar b$ separated by $r$. Eigen-energies are presented by symbols in Figs. \ref{fig:2} and \ref{fig:3} for all four sets of quantum numbers shown in Table \ref{tab:1}. The colors of symbols indicate which Fock component dominates an eigenstate, as determined from the normalized overlaps of an eigenstate $|n\rangle$ to operators $O_i$. For more details on the calculation of the eigen-energies and the overlaps see Sec. IV and Appendix in \cite{Sadl:2021bme}.

The lines in Figs. \ref{fig:2} and \ref{fig:3} provide the related non-interacting (n. i.) energies $E_n$ of two-hadron states
\begin{equation}
\begin{split}
  E^{\textrm{n.\,i.}}_{B\bar B^*}\!&=\!2m_B,  \qquad\qquad\quad E^{\textrm{n.\,i.}}_{[\bar b(0) b(r)] l(0)}\!=\!V_{\bar bb}(r)+m_{l}\>,\\    
  E^{\textrm{n.\,i.}}_{[\bar b(0) b(r)]l(\vec p)}\!&=\!V_{\bar bb}(r)+E_{l(\vec p)},\quad l=\pi, \rho, b_1, a_1, a_0\>,
\label{E6}
\end{split}
\end{equation}
where $\bar bb$ static potential $V_{\bar bb}(r)$,   $m_{l}$ and $m_B=m_{B^*}=0.5201(19)$ (mass of $B^{(*)}$ for $m_b\to \infty$ without $b$ rest mass) are determined on the same lattice. $E_{l(\vec p)}$ is determined using $[\bar q\Gamma^\prime q]_{\vec p}$ and approximately satisfies $E_{l(\vec p)}\simeq \sqrt{m_l^2+\vec p^2}$ (see the last paragraph in Sec. \ref{sec:QNandOP}).
   
All observed eigen-energies $E_n$  of the $\bar bb\bar du$ system (symbols) are very close to non-interacting energies $E^{n.i.}$ of $[\bar bb][\bar du]$ or $[\bar bu][\bar db]$ (lines). This represents the most important conclusion of the present study. In particular, eigenstates dominated by  $[\bar bb][\bar du]$ operators have energies consistent with the sum of energies for $[\bar bb(r)]$ and $[\bar du]$. Given our precision, we therefore do not observe attraction or repulsion between bottomonium and light hadrons for the considered separations $r$.

\begin{figure*}[ht!]
\centering
\begin{subfigure}{0.41\textwidth}
\includegraphics[width=\textwidth]{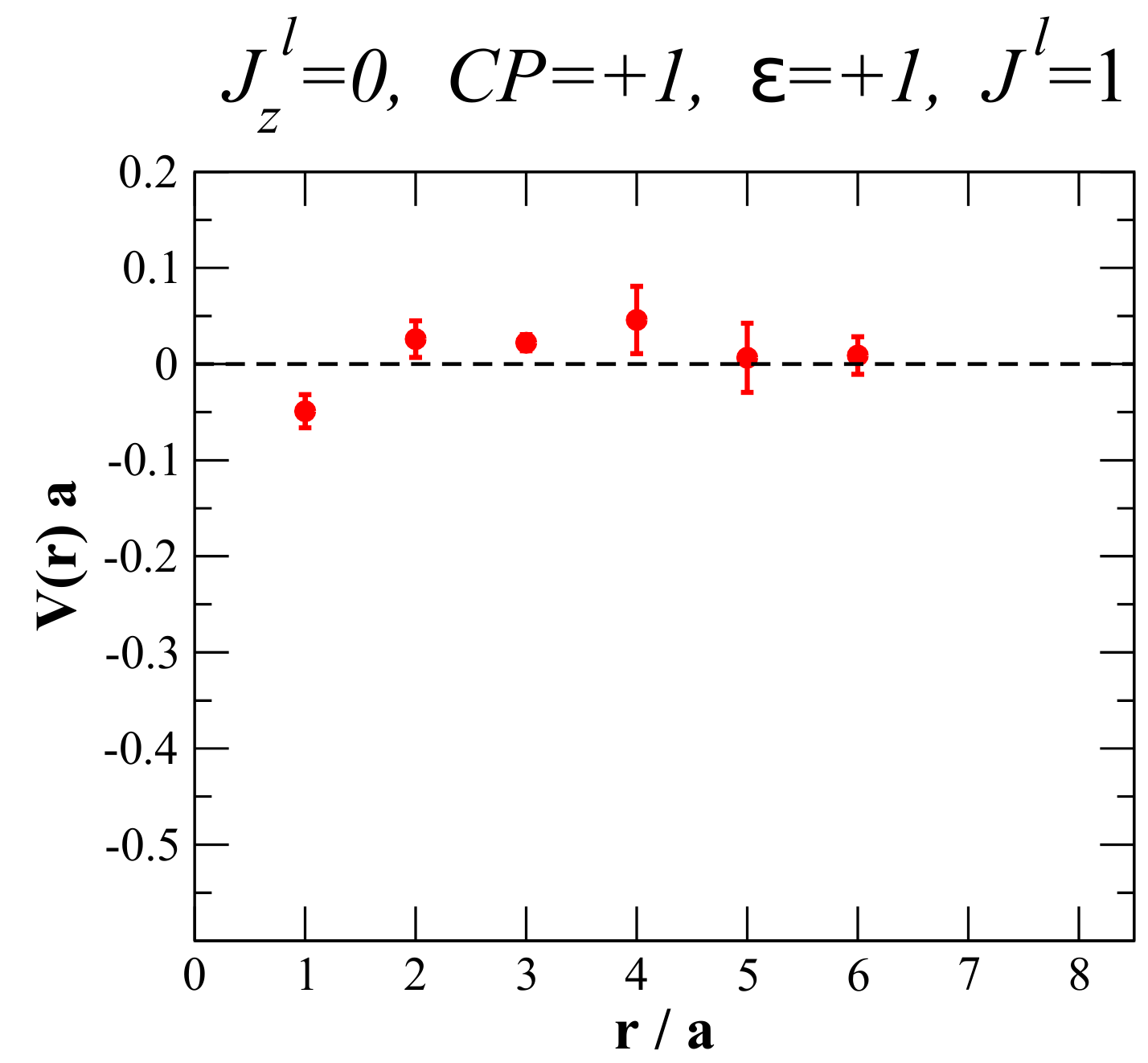}
\caption{ } \label{fig:5a}
\end{subfigure} 
\hspace*{1cm}
\begin{subfigure}{0.41\textwidth}
\includegraphics[width=\textwidth]{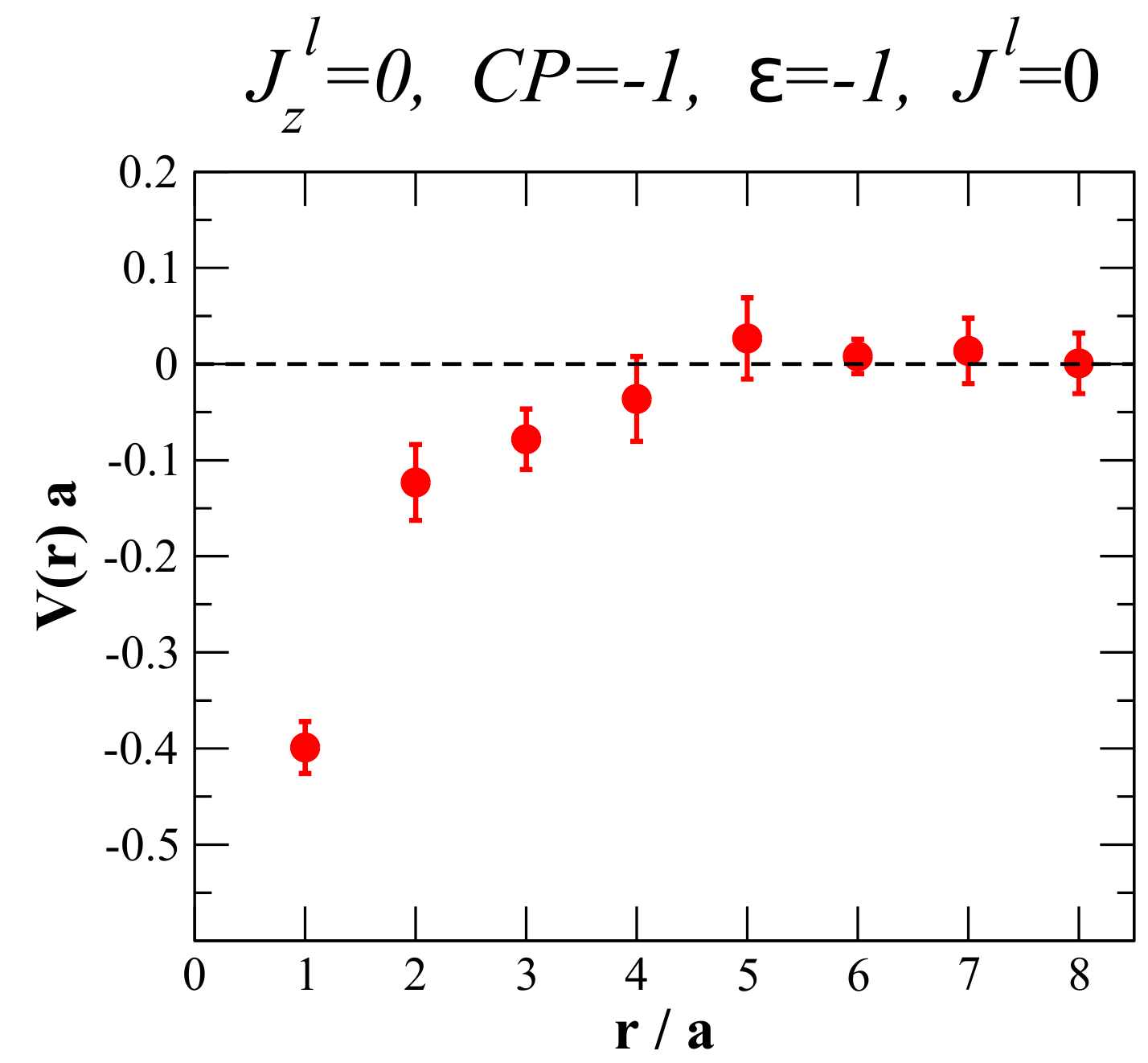}
\caption{ } \label{fig:5b}
\end{subfigure}
\caption{Static potentials between $B$ and $\bar  B^{*}$ separated by $r$ from lattice simulations (see Fig. \ref{fig:1a}).  Quantum numbers $\mathrm{(a)}\  I=1,~J^l=1,~J^l_z=0,~C\!\cdot\! P\!=\! \epsilon\!=\!+1$ are considered here and (b)  $I=1,~J^l=0,~J^l_z=0,~C\!\cdot\! P\!=\! \epsilon\!=\!-1$ were studied in \cite{Prelovsek:2019ywc}. The potential (a) is consistent with zero for $r/a\geq 2$ within slightly more than one sigma errors, which are shown in the plot. Both simulations are performed on the same ensemble with the lattice spacing $a\simeq\SI{0.124}{fm}$.}
\label{fig:5}
\end{figure*}

For the three quantum channels we consider in this study, the eigenstate dominated by $B\bar B^*$ is present only in $C\!\cdot\! P\!=\!\epsilon\!=\!+1$. Its energy $E_{B\bar B^*}(r)$ is represented by the red circles in Fig. \ref{fig:2a} and is close to $m_B+m_{B^*}$. For the reminder of the discussion, we assume that this eigenstate couples only to $B\bar B^*$ Fock component and does not contain other Fock components, which is supported by the extracted normalized overlaps. The energy of this eigenstate represents the total energy without the kinetic energy of heavy degrees of freedom. The difference $V (r) = E_{B\bar B^*}(r)-m_B-m_{B^*}$, therefore, represents the potential felt by the heavy degrees of freedom, in this case between $B$ and $\bar B^*$. Possible implications of the potentials in Fig. \ref{fig:5} for $Z_b$ are discussed below.
\begin{itemize}
\item   $J^l\!=\! 0~ \&~ C\!\cdot\! P\!=\! \epsilon\!=\!-1$: The  potential with sizable attraction between $B$ and $\bar B^{*}$ at small $r$ has been found  \cite{Prelovsek:2019ywc,Peters:2016wjm}. The results from \cite{Prelovsek:2019ywc}, which are obtained on the same  ensemble as employed here, are shown in Figs. \ref{fig:2b} and \ref{fig:5b}. Motion of $B$ and $\bar B^*$ with experimental masses in this potential leads to one $B\bar B^*$ bound state below  threshold, whose binding energy depends on the parametrization of the potential.  Assuming the non-singular potential  $V(r)=-A r^{-(r/d)^F}$ lead to the range of binding energies $M-m_B-m_{B^*}=-48 ^{+41}_{-{108}}\,$MeV \cite{Prelovsek:2019ywc}. Some parametrizations among those lead to a bound state closely below threshold ($\simeq \SI{20}{MeV}$) and sharp peak in the $B\bar B^*$ rate above threshold $-$ a feature that could be related to the observed experimental $Z_b$  peak. Most of the parametrizations in \cite{Prelovsek:2019ywc} lead to a binding energy larger than $\SI{20}{MeV}$ and a less significant peak in the rate above threshold, since the size of the peak decreases as the binding energy increases. The singular form of the potential $V(r)=-\tfrac{A}{r} r^{-(r/d)^F}$ would also  lead  to one bound state, but with a  larger binding energy. This component is therefore significantly attractive: it is possible that this component alone is to attractive and  leads to a binding energy that is to large  in comparison with experimental $Z_b$. 
\item $ J^l=1~ \&~ C\!\cdot\! P\!=\! \epsilon\!=\!+1$: The potential for this component in Fig. \ref{fig:5a} shows no observable attraction or repulsion between $B$ and $\bar B^{*}$   at $r\geq \SI{0.2}{fm}$ and a very mild attraction at $r\simeq \SI{0.1}{fm}$.
\item Linear combination: The $Z_b$ is a linear combination of those two quantum numbers \eqref{decomposition}. The $B\bar B^*$ and $ B^*\bar B^*$ channels are coupled in this system  via the  strongly attractive potential  for component  $C\!\cdot\! P\!=\! \epsilon\!=\!-1$ and very mildly attractive potential for $C\!\cdot\! P\!=\! \epsilon\!=\!+1$, both shown in Fig. \ref{fig:5}.  It is not possible to  establish implications concerning $Z_b$ at present since  neither of these potentials is known  from the lattice simulations in detail.  However, it is conceivable that a mutual effect of a significantly attractive and a very mildly attractive potential could lead to a bound state closely below $B\bar B^*$ threshold, which could be related to experimental $Z_b$.
\end{itemize}
Let us note that the $Z_b(10610)$ was found as a virtual bound state  slightly below threshold by the re-analysis of the experimental data \cite{Wang:2018jlv}.

\section{Outlook}\label{sec:outlook}

The presented simulations of $\bar bb\bar du$  system represent only the first step towards exploring  the energy region near $m_{Zb}\simeq m_{B}+m_{B^*}$, where a number of severe simplifications have been made. It would be valuable if the future lattice simulation could determine the eigen-energies of the considered channels with  an improved accuracy. The simulations with smaller lattice spacing would be needed to extract static potentials at smaller separations between static quarks. The simulations with larger volumes would be more challenging since the discrete spectrum of $[\bar bb]l(p)$ states would be denser. A much more severe challenge would be to take into account the resonance nature of  $\rho,~b_1,~a_1,~a_0$ decaying to multiple hadrons, which will require implementation of multi-hadron operators $O_{[\bar bb] l_1(p_1) l_2(p_2) ... }$. Furthermore, an analytic study that considers the dynamics of the  $B\bar B^*$ and $B^*\bar B^*$ channels  which are coupled via the potentials in Fig. \ref{fig:5} would be needed.

\section{Conclusions}\label{sec:conclusions}

Two $Z_b$ resonances with $J^P=1^+$  were the first discovered bottomonium-like tetraquarks. They predominantly decay to $B\bar B^*$ and $B^*\bar B^*$ and lie slightly above these two thresholds. They decay also to a bottomonium and a pion, which implies the exotic quark content  $\bar bb\bar du$.  Our aim  is to explore whether the interaction between $B^{(*)}$ and $\bar B^*$ is responsible for the existence of $Z_b$. The main challenge is that  $Z_b$ decays to $\bar bu+\bar du$ as well as lower lying states $\bar bb+\bar du$

We study the system $\bar bb\bar du$ with  static $\bar{b}b$ pair separated by $r$ on the lattice. Four quantum channels are considered and operators of type $[\bar bu][\bar du]$ and $[\bar bb][\bar du]$ are employed. We determine eigen-energies $E_n(r)$ and compare them to the non-interacting energies of two-hadron systems.

The  $Z_b$ with finite $m_b$ can decay to $\Upsilon \pi$ and $\eta_b \rho$ (among others), while these two quantum channels are decoupled for the static $b$ quarks used in the simulation. The simulation \cite{Prelovsek:2019ywc} considered the quantum number that couples to $\Upsilon \pi$ and found that the potential between $B$ and $\bar B^*$ is significantly attractive at $r<\SI{0.4}{fm}$. The present simulation considers the quantum number that couples to $\eta_c \rho$ and finds that the potential between $B$ and $\bar B^*$ is consistent with zero, except for a slight attraction at $r\simeq 0.1~$fm. The  first attractive potential alone leads to a bound state below $m_B+m_{B^*}$ that could be related to $Z_b$ \cite{Prelovsek:2019ywc}, but it is likely somewhat to deep. It is conceivable that the mutual effect of both potentials could lead to a $Z_b$ state in the vicinity of the $m_B+m_{B^*}$ threshold. The conclusion is also that the interaction between bottomonium and light hadrons for all four explored quantum channels $\bar bb\bar du$ is small.

\acknowledgments
We thank  V. Baru, P. Bicudo, N. Brambilla,  T. Cohen, C. Hanhart, M. Karliner, R. Mizuk, J. Soto, A. Peters, J. Tarrus and M. Wagner for valuable discussions.
S.P. acknowledges support by Slovenian Research Agency ARRS (research core funding No. P1-0035 and No. J1-8137) and DFG grant No. SFB/TRR 55. The work of M. S. is supported by Slovenian Research Agency ARRS (Grant No. 53647).


\providecommand{\href}[2]{#2}\begingroup\raggedright\begin{thebibliography}{10}

\bibitem{Belle:2011aa}
{\scshape Belle} collaboration, \emph{{Observation of two charged
  bottomonium-like resonances in Y(5S) decays}},
  \href{https://doi.org/10.1103/PhysRevLett.108.122001}{\emph{Phys. Rev. Lett.}
  {\bfseries 108} (2012) 122001}
  [\href{https://arxiv.org/abs/1110.2251}{{\ttfamily 1110.2251}}].

\bibitem{Garmash:2014dhx}
{\scshape Belle} collaboration, \emph{{Amplitude analysis of $e^+e^- \to
  \Upsilon(nS) \pi^+\pi^-$ at $\sqrt{s}=10.865$~GeV}},
  \href{https://doi.org/10.1103/PhysRevD.91.072003}{\emph{Phys. Rev.}
  {\bfseries D91} (2015) 072003}
  [\href{https://arxiv.org/abs/1403.0992}{{\ttfamily 1403.0992}}].

\bibitem{ParticleDataGroup:2020ssz}
{\scshape Particle Data Group} collaboration, \emph{{Review of Particle
  Physics}}, \href{https://doi.org/10.1093/ptep/ptaa104}{\emph{PTEP} {\bfseries
  2020} (2020) 083C01}.

\bibitem{Wang:2018jlv}
Q.~Wang, V.~Baru, A.A.~Filin, C.~Hanhart, A.V.~Nefediev and J.L.~Wynen,
  \emph{{Line shapes of the $Z_b(10610)$ and $Z_b(10650)$ in the elastic and
  inelastic channels revisited}},
  \href{https://doi.org/10.1103/PhysRevD.98.074023}{\emph{Phys. Rev.}
  {\bfseries D98} (2018) 074023}
  [\href{https://arxiv.org/abs/1805.07453}{{\ttfamily 1805.07453}}].

\bibitem{Bondar:2011ev}
A.E.~Bondar, A.~Garmash, A.I.~Milstein, R.~Mizuk and M.B.~Voloshin,
  \emph{{Heavy quark spin structure in $Z_b$ resonances}},
  \href{https://doi.org/10.1103/PhysRevD.84.054010}{\emph{Phys. Rev. D}
  {\bfseries 84} (2011) 054010}
  [\href{https://arxiv.org/abs/1105.4473}{{\ttfamily 1105.4473}}].

\bibitem{BO}
M.~Born and J.~Oppenheimer, \emph{{ On the quantum theory of Molecules}},
  \href{https://doi.org/10.1103/PhysRevD.78.054511}{\emph{Ann. Physik}
  {\bfseries 84} (1927) 457}.

\bibitem{Braaten:2014qka}
E.~Braaten, C.~Langmack and D.H.~Smith, \emph{{Born-Oppenheimer Approximation
  for the XYZ Mesons}},
  \href{https://doi.org/10.1103/PhysRevD.90.014044}{\emph{Phys. Rev.}
  {\bfseries D90} (2014) 014044}
  [\href{https://arxiv.org/abs/1402.0438}{{\ttfamily 1402.0438}}].

\bibitem{Sadl:2021bme}
M.~Sadl and S.~Prelovsek, \emph{{Tetraquark systems $\bar bb\bar du$ in static
  limit and lattice QCD}},  \href{https://arxiv.org/abs/2109.08560}{{\ttfamily
  2109.08560}}.

\bibitem{Peters:2016wjm}
A.~Peters, P.~Bicudo, K.~Cichy and M.~Wagner, \emph{{Investigation of $B\bar B$
  four-quark systems using lattice QCD}},
  \href{https://doi.org/10.1088/1742-6596/742/1/012006}{\emph{J. Phys. Conf.
  Ser.} {\bfseries 742} (2016) 012006}
  [\href{https://arxiv.org/abs/1602.07621}{{\ttfamily 1602.07621}}].

\bibitem{Prelovsek:2019ywc}
S.~Prelovsek, H.~Bahtiyar and J.~Petkovic, \emph{{Zb tetraquark channel from
  lattice QCD and Born-Oppenheimer approximation}},
  \href{https://doi.org/10.1016/j.physletb.2020.135467}{\emph{Phys. Lett. B}
  {\bfseries 805} (2020) 135467}
  [\href{https://arxiv.org/abs/1912.02656}{{\ttfamily 1912.02656}}].

\bibitem{Alberti:2016dru}
M.~Alberti, G.S.~Bali, S.~Collins, F.~Knechtli, G.~Moir and W.~S\"oldner,
  \emph{{Hadroquarkonium from lattice QCD}},
  \href{https://doi.org/10.1103/PhysRevD.95.074501}{\emph{Phys. Rev. D}
  {\bfseries 95} (2017) 074501}
  [\href{https://arxiv.org/abs/1608.06537}{{\ttfamily 1608.06537}}].

\bibitem{Juge:1999ie}
K.~Juge, J.~Kuti and C.~Morningstar, \emph{{Ab initio study of hybrid anti-b g
  b mesons}}, \href{https://doi.org/10.1103/PhysRevLett.82.4400}{\emph{Phys.\
  Rev.\ Lett.} {\bfseries 82} (1999) 4400}
  [\href{https://arxiv.org/abs/hep-ph/9902336}{{\ttfamily hep-ph/9902336}}].

\bibitem{Hasenfratz:2008ce}
A.~Hasenfratz, R.~Hoffmann and S.~Schaefer, \emph{{Low energy chiral constants
  from epsilon-regime simulations with improved Wilson fermions}},
  \href{https://doi.org/10.1103/PhysRevD.78.054511}{\emph{Phys. Rev.}
  {\bfseries D78} (2008) 054511}
  [\href{https://arxiv.org/abs/0806.4586}{{\ttfamily 0806.4586}}].

\bibitem{Lang:2011mn}
C.B.~Lang, D.~Mohler, S.~Prelovsek and M.~Vidmar, \emph{{Coupled channel
  analysis of the rho meson decay in lattice QCD}},
  \href{https://doi.org/10.1103/PhysRevD.89.059903,
  10.1103/PhysRevD.84.054503}{\emph{Phys. Rev.} {\bfseries D84} (2011) 054503}
  [\href{https://arxiv.org/abs/1105.5636}{{\ttfamily 1105.5636}}].

\end{thebibliography}\endgroup

\end{document}